\documentclass{moriond}
\usepackage{subcaption}
\newcommand{\Photo}{}

\def\Journal#1#2#3#4{{#1} {\bf #2}, #3 (#4)}

\def\PRL{\em Phys. Rev. Lett.}

\def\be{\begin{equation}}
\def\ee{\end{equation}}
\def\bea{\begin{eqnarray}}
\def\eea{\end{eqnarray}}

\usepackage{subcaption}
\usepackage{graphicx}

\begin{document}
\vspace*{4cm}
\title{KM3-230213A and potential astrophysical sources}

\author{P.A. Sevle Myhr \\ On behalf of the KM3NeT Collaboration}
\address{CP3, Chem. du Cyclotron 2, 1348 Louvain-la-Neuve}

\maketitle
\abstract{The recent detection of the ultra-high energy neutrino KM3-230213A by KM3NeT/ARCA marks the first observation of an astrophysical neutrino with energy above 100 PeV, opening a new window to the ultra-high energy Universe. In this contribution, the current global ultra-high energy neutrino landscape in light of this event is reviewed, including tension of this observation with existing limits set by the IceCube and the Pierre Auger Observatories. Different scenarios are discussed to explain its origin. Recent efforts to constrain features of potential source populations using the inferred diffuse ultra-high energy neutrino flux are also presented.}

\section{Introduction}
Over the past 15 years, neutrino astronomy has developed into a mature field with accumulating evidence for multiple cosmic accelerators~\cite{IceCube:2018dnn}$^,$ \cite{IceCube:2022der}. In addition to evidence of neutrino excess from individual sources, details in the spectrum of astrophysical neutrinos, extending over many decades in energy have been studied~\cite{IceCube:2025tgp}. Despite the discoveries in the TeV--PeV energy range, ultra-high energy (UHE) neutrinos, with energy greater than 50 PeV, have remained elusive until the detection of KM3-230213A.

Located about 100 km off the south-eastern coast of Sicily, Italy, KM3NeT/ARCA is a cubic-kilometre neutrino telescope currently under construction and taking data in partial configurations since the end of 2015~\cite{KM3Net:2016zxf}. The detector consists of an array of digital optical modules (DOMs), each equipped with 31 photomultipliers and arranged in lines of 18 DOMs anchored to the bottom of the Mediterranean Sea. These detectors record the Cherenkov photons produced when charged particles propagate in the sea water faster than light. On February 13th 2023, the 21-line configuration of the KM3NeT/ARCA detector recorded an almost horizontal muon track with a reconstructed energy of $120^{+110}_{-60}$ PeV~\cite{KM3NeT:2025npi}. The total amount of rock and sea water along the direction of the muon strongly disfavour the hypothesis of an atmospheric origin, leaving an astrophysical neutrino as the most likely explanation. Assuming the astrophysical neutrino spectrum at this energy follows an E$^{-2}$ power-law, the energy of the initial neutrino is estimated around $220^{+570}_{-110}$ PeV. This event marks the first detection of astrophysical neutrinos in the UHE regime and provides the first direct opportunity to study UHE neutrinos and their connection to ultra-high-energy cosmic rays (UHECRs).

\section{The global ultra-high energy neutrino landscape}
The detection of KM3-230213A is in tension with the non-observation of neutrinos in the same energy range by the IceCube and Pierre Auger observatories, both of which have higher exposures than KM3NeT/ARCA~\cite{KM3NeT:2025cpp}. The tension was found to be about 2.5$\sigma$ with a joint-fit, single-flavour UHE neutrino flux normalisation of $E^2\Phi=7.5^{+13.1}_{-4.7}\times10^{-10}$ GeV~cm$^{-2}$~s$^{-1}$~sr$^{-1}$. In addition, a joint-fit with IceCube diffuse measurements at lower energies found no preference for an additional component to the diffuse astrophysical neutrino spectrum at higher energies, and the diffuse flux of astrophysical neutrinos remains consistent with a power-law from TeV to hundreds of PeV.

\section{Candidate sources}
\subsection{Steady sources}
A natural question to ask after the detection of KM3-230213A is what types of astrophysical environments can accelerate particles required to produce UHE neutrinos, and whether any known sources in the arrival direction of KM3-230213A can be associated to this event. Potential galactic counterparts were considered in Ref~\cite{KM3NeT:2025pre}, but the lack of powerful, Galactic emitters within the uncertainty region of the direction of the event, combined with the lack of gamma-ray sources found by either HAWC or LHAASO rules out the galaxy as a source for KM3-230213A.

Within the $\sim$$3^\circ$ uncertainty region around the direction of the event, 17 blazars has been identified and three of these were flaring at different wavelengths around the time of KM3-230213A~\cite{KM3NeT:2025bxl}. The flaring blazars are MRC 0614-083, with an increase in the X-ray light curve multiple years in advance of KM3-230213A, the bright blazar PKS 0605-085 with a flare in the \textit{Fermi}-LAT light curve peaking around 181 days before KM3-230213A, and PMN J0606-0724, which exhibit a strong 15 GHz radio-flare peaking just 5 days before KM3-230213A. Out of the 17 identified blazars, no conclusive association was found with the UHE event.

\begin{figure}
    \centering
    \begin{subfigure}[t]{0.45\textwidth}
        \centering
        \includegraphics[width=\linewidth]{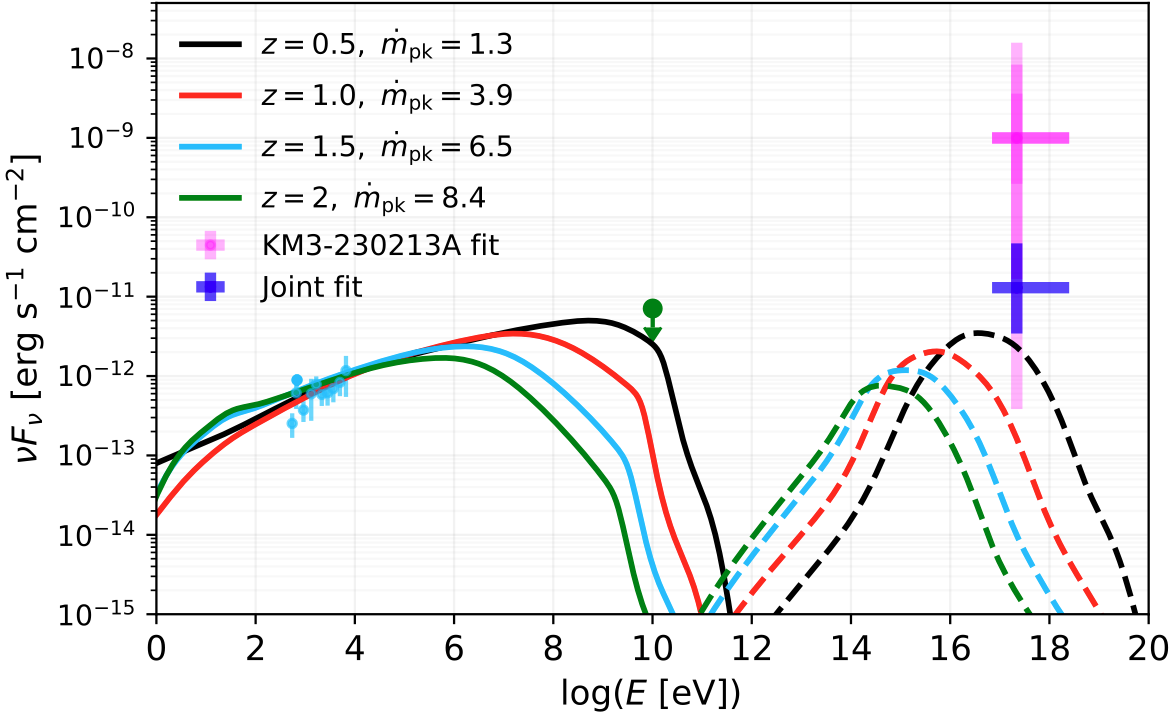} 
    \end{subfigure}
    \hfill
    \begin{subfigure}[t]{0.54\textwidth}
        \centering
        \includegraphics[width=\linewidth]{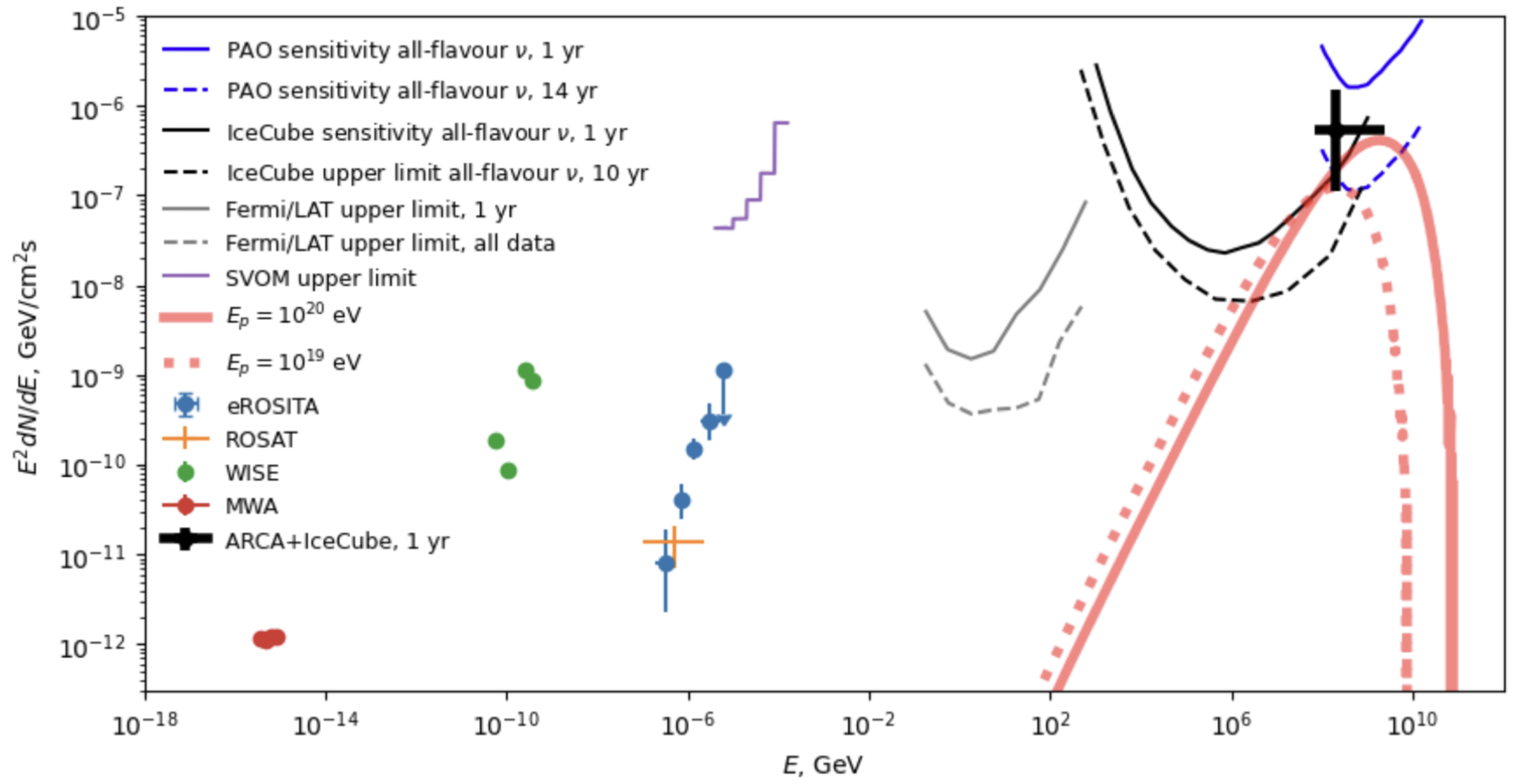} 
    \end{subfigure}
    \hfill
    \caption{\textit{Left:} Spectral fits for gamma-ray emission and UHE neutrino emission from Ref~\protect\cite{Yuan:2025zwe}. Different colour lines correspond to different source redshift and peak accretion rate. \textit{Right:} Sensitivity, limits and measurements from different experiments as detailed in Ref.~\protect\cite{Neronov:2025jfj}. The solid (dotted) pink lines correspond to the predicted UHE neutrino flux from a mono-energetic proton population with $10^{20}$ ($10^{19}$) eV energy.}
    \label{fig:source}
\end{figure}

\subsection{Transient sources}
From the aforementioned flaring blazars, PMN J0606-0724 was further investigated independently by Ref.~\cite{deClairfontaine:2025gei} and~\cite{Kivokurtseva:2025sui}. In Ref.~\cite{deClairfontaine:2025gei}, the authors argue that a red giant star crossing the blazar jet could act as both a source of additional baryons, allowing the baryon energy density to temporarily surpass the magnetic energy density, and provide an additional photon field leading to increased photo-hadronic interactions capable of producing neutrinos at PeV energies. The authors of Ref.~\cite{Kivokurtseva:2025sui} demonstrated that increased activity in the bright radio-core of the jet can sufficiently explain the UHE neutrino, as the required source proton power would need to be comparable to the observed photon luminosity. In both cases, the corresponding high-energy gamma-rays would cascade down to lower energy radiation through pair production, making it consistent with the lack of gamma-ray observations from PMN J0606-0724.

In Ref.~\cite{Yuan:2025zwe}, the authors performed detailed multi-messenger emission modelling of the X-ray flaring blazar MRC 0614-083. Combining infrared, optical and X-ray light curves, they showed that a period of super-Eddington accretion can provide sufficient photon fields for photo-hadronic interactions to achieve a neutrino flux compatible with the joint-fit flux estimate of KM3-230213A, whilst simultaneously explaining the associated X-ray flare by internal electromagnetic cascades. As the redshift of MRC 0614-083 is not know, they provide flux predictions from a few different scenarios in the left panel of Figure~\ref{fig:source}.

Instead of considering a single astrophysical source in detail, qualitative statements can be made on isolated sources with flaring activity. In Ref.~\cite{Neronov:2025jfj}, the authors considered the general features required for a transient source to produce a KM3-230213A-like event. They found that the source must have a hard proton spectrum to be consistent with IceCube sensitivities to year-long transients at ultra-high energies, and found that this can be achieved with a mono-energetic distributions of protons as shown in the right panel of Figure~\ref{fig:source}. In addition, they show that the non-observations of UHE neutrinos by IceCube constrains the total rate of such transient sources across the full sky to be $\leq 0.4$ per year.

\subsection{Cosmogenic origin}
Cosmogenic neutrinos -- created by interaction of UHECRs with the cosmic microwave background and other extragalactic background photon fields during propagation -- are expected to make up a significant fraction of the UHE neutrino flux based on detection of cosmic rays at the highest energies. For the UHE neutrino to be cosmogenic, a population of parent cosmic-ray sources needs to be strongly positively evolving with an evolution rate comparable to or surpassing the star formation rate~\cite{KM3NeT:2025aps}$^,$~\cite{Cermenati:2025ogl}. In addition, the parent cosmic ray population needs to be quite hard, with a significant UHECR proton fraction at the highest energies. In Ref.~\cite{Alhebsi:2026bdk}, the authors consider a two-population model for UHECRs comprising a mixed composition at intermediate-to-high energies, together with a subdominant pure-proton component characterised by higher average rigidity in line with the measured UHECR spectrum. Including the null observations from IceCube and Pierre Auger, the hypothesis suggested by the authors disfavours a similarly strong source evolution whilst respecting the measured diffuse GeV-TeV gamma-ray flux.

A different scenario was investigated in Ref~\cite{Das:2025vqd} where the authors considered line-of-sight interactions originating from a highly collimated UHECR source in the direction of KM3-230213A. Due to the deflection of extragalactic magnetic fields, they disfavour the UHE neutrino flux associated with KM3-230213A only due to unrealistic high source proton luminosity. However, the UHE neutrino flux associated with the KM3NeT, IceCube and Auger joint-fit can be reached in this scenario by a powerful blazar with synchrotron luminosity comparable to the expected source UHECR luminosity, or by energetic transients like long-duration GRBs.

\section{Constraining source population}
\begin{figure}[t]
    \centering
    \begin{subfigure}[t]{0.3\textwidth}
        \centering
        \includegraphics[width=\linewidth]{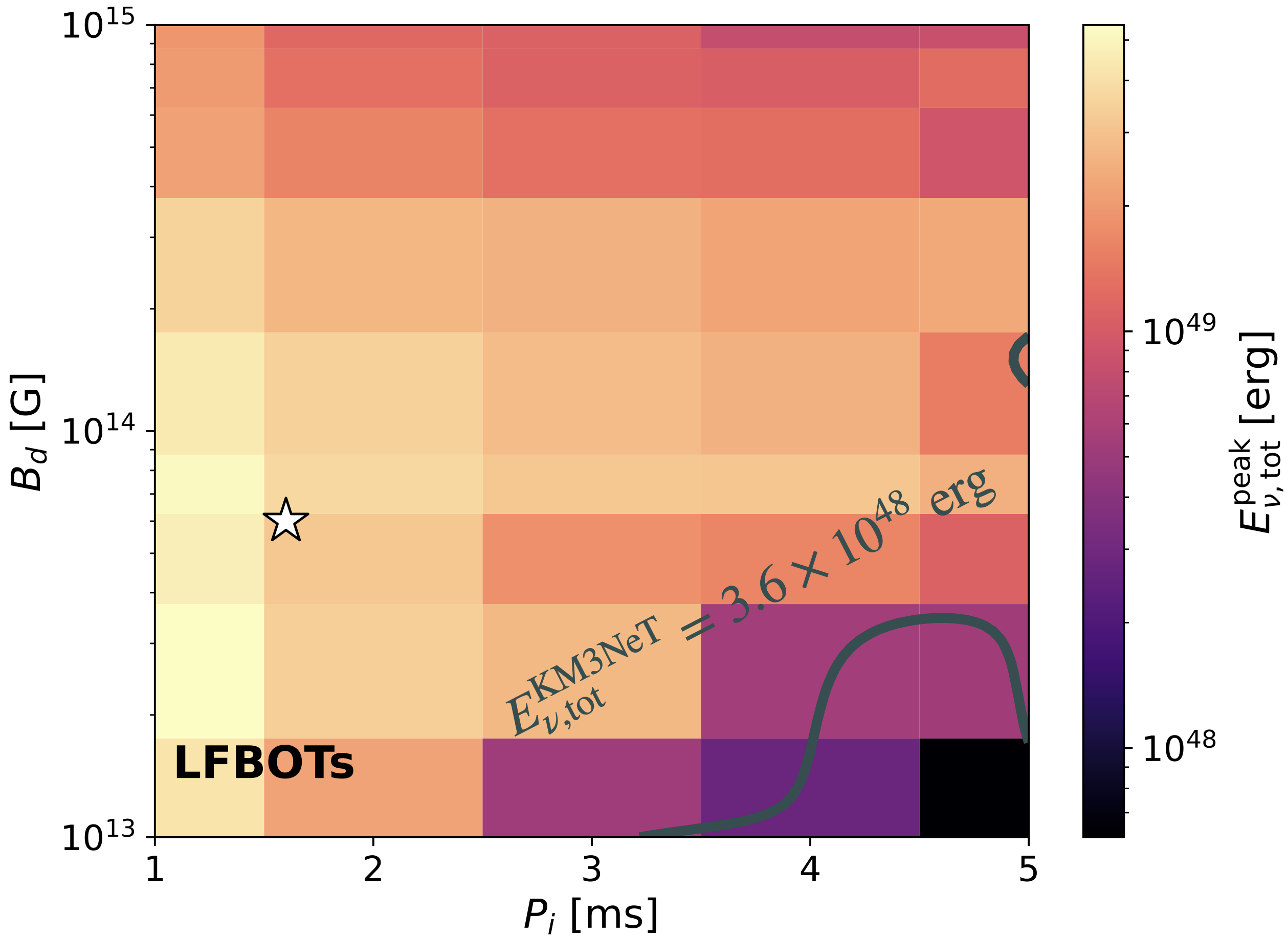} 
    \end{subfigure}
    \begin{subfigure}[t]{0.3\textwidth}
        \centering
        \includegraphics[width=\linewidth]{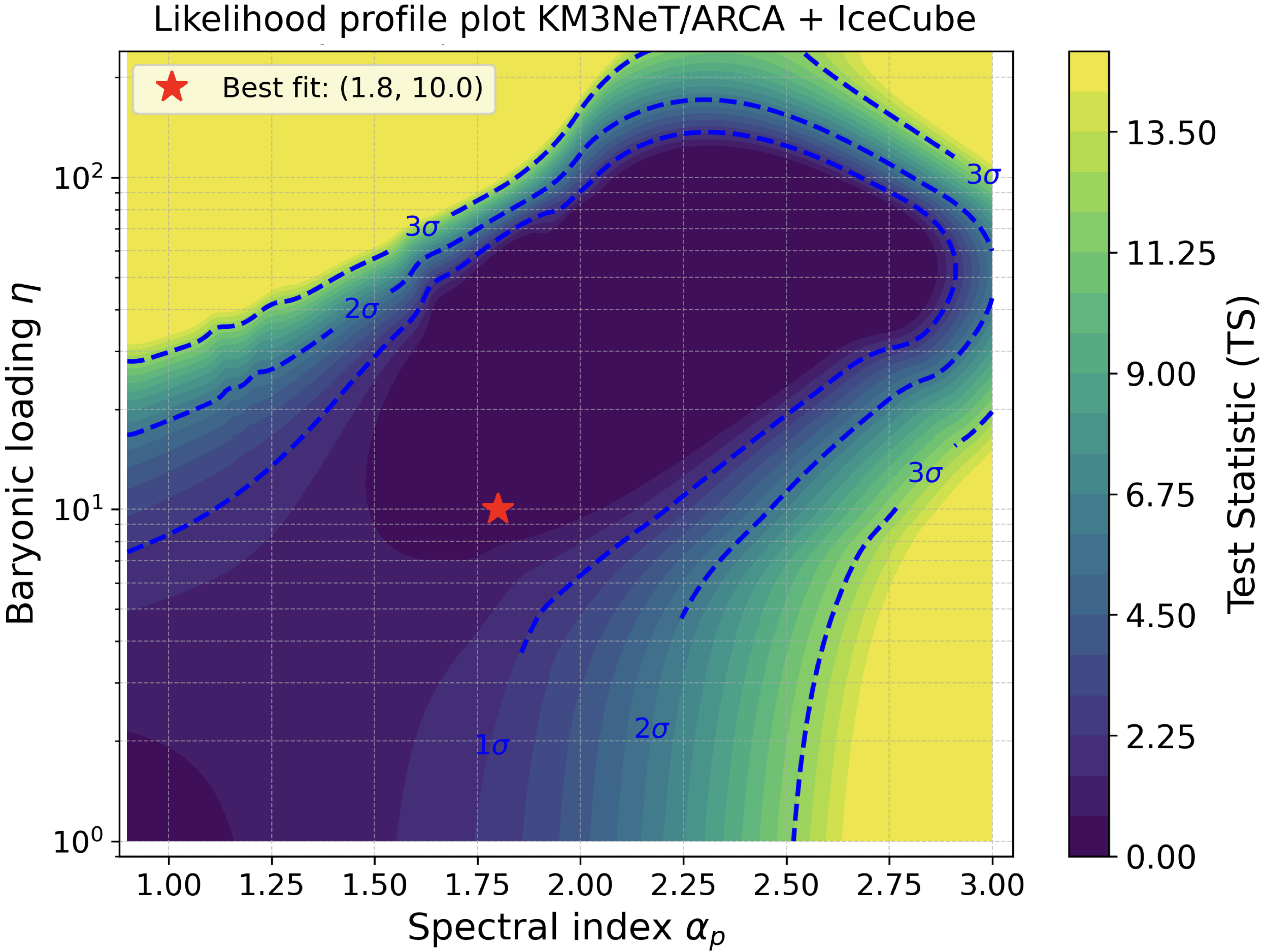} 
    \end{subfigure}
    \begin{subfigure}[t]{0.3\textwidth}
    \centering
        \includegraphics[width=\linewidth]{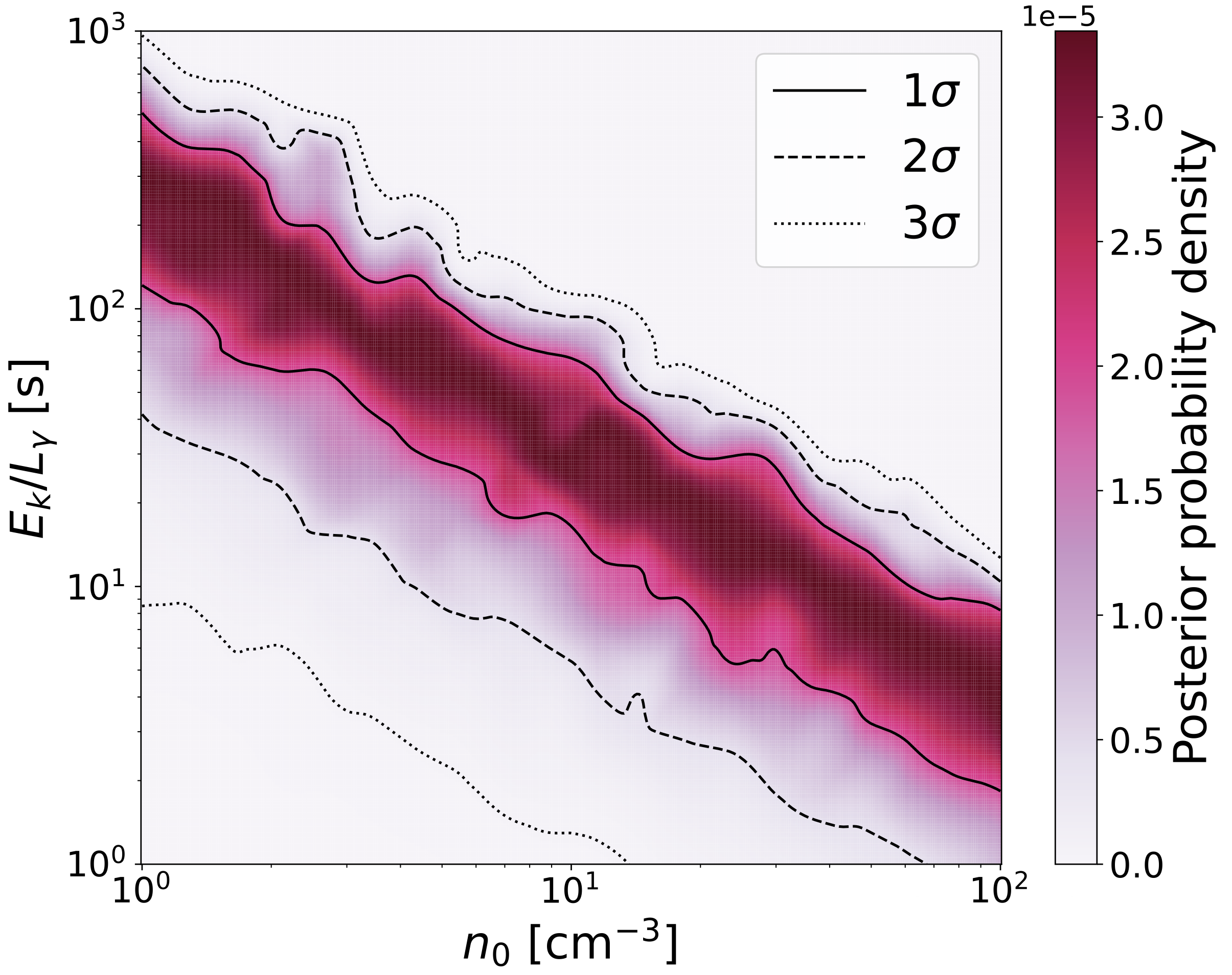} 
    \end{subfigure}
    \caption{\textit{Left:} Parameter space scan of dipolar magnetic field strength and initial spin period of the luminous fast blue optical transient population considered in Ref.~\protect\cite{Mukhopadhyay:2026wrv}. \textit{Middle:} Likelihood map of the parameter space scan for baryon loading and proton spectral index for a population of blazars~\protect\cite{KM3NeT:2025lly}. \textit{Right:} Posterior probability density of the parameter space scan in ratio of kinetic energy to gamma-ray luminosity and interstellar medium density for the long-duration GRB population considered in Ref.\protect\cite{KM3NeT:2025zmb}.}
    \label{fig:scan}
\end{figure}

Under the hypothesis of an all-sky diffuse astrophysical UHE neutrino flux, the first positive observation provided by KM3-230213A can be used to constrain astrophysical source populations. A population of pulsar-powered optical transient was considered in Ref.~\cite{Mukhopadhyay:2026wrv} where the authors considered three different populations; supernovae, super-luminous supernovae and luminous fast blue optical transients. For each source population, they perform a parameter space scan in dipolar magnetic field strength and initial spin period shown in Figure~\ref{fig:scan} (left). They found the best candidate to be luminous fast blue optical transients which reach the required diffuse neutrino fluence associated with the UHE joint-fit, but the peak of the emission occurs at lower energies.

In a similar manner, the KM3NeT Collaboration considered a population of blazars by modelling the combined gamma-ray and UHE neutrino fluxes~\cite{KM3NeT:2025lly}. By performing a parameter space scan in baryon loading and proton spectral index, they found that a population of blazars can reach sufficient UHE neutrino fluence whilst respecting the diffuse gamma-ray spectrum measured by \textit{Fermi}-LAT. They find a best-fit baryon loading of $\sim10$ and spectral proton index $\sim 1.8$, shown in Figure~\ref{fig:scan} (middle), to explain the joint-fit flux from a population of blazars.

Finally, the KM3NeT Collaboration have considered a population of long-duration GRBs as potential source of the UHE neutrino~\cite{KM3NeT:2025zmb}. In this work, a population of GRBs blastwaves were combined with the corresponding \textit{Swift}- and \textit{Fermi}-GBM-inferred luminosity functions to perform a parameter space scan in baryon loading and density of the surrounding medium. This scan in shown in Figure~\ref{fig:scan} (right). The baryon loading is constrained to be $\leq 392$ at 90\% confidence for interactions with a standard interstellar medium density.

\section{Outlook}
The first detection of an ultra-high energy neutrino marks a major step toward understanding the UHE neutrino sky and its connection to UHECR accelerators. Alongside conventional astrophysical sources, a range of more exotic scenarios has been proposed, including the decay of dark matter and radiation from primordial black hole evaporation. The continued expansion of the KM3NeT/ARCA detector toward its final gigaton-scale configuration will be essential for probing the origin of KM3-230213A-like events and further constraining the UHE neutrino flux.

\end{document}